\documentclass[a4paper,apl,reprint,floatfix]{revtex4-1}

\usepackage{subfigure}
\usepackage{hyperref}
\usepackage{ifpdf}
\ifpdf
\usepackage{graphicx}
\usepackage{epstopdf}
\AppendGraphicsExtensions{.tif}
\else
\usepackage{graphicx}
\fi
\pdfoutput=1

\begin{document}

\title{Single-hole tunneling through a two-dimensional hole gas in intrinsic silicon.} 
\author{Paul C. \surname{Spruijtenburg}}
\email[To whom correspondence should be addressed: ]{p.c.spruijtenburg@utwente.nl}

\author{Joost \surname{Ridderbos}}
\author{Filipp \surname{Mueller}}
\author{Anne W. \surname{Leenstra}}
\author{Matthias \surname{Brauns}}
\author{Antonius A.I. \surname{Aarnink}}
\author{Wilfred G. \surname{van der Wiel}}
\author{Floris A. \surname{Zwanenburg}}
\affiliation{NanoElectronics Group, MESA+ Institute for Nanotechnology, University of Twente, Enschede The Netherlands}
\keywords{silicon; quantum dot; holes; single-charge tunneling; two-dimensional hole gas; single-hole transistor}

\date{\today}

\begin{abstract}
In this letter we report single-hole tunneling through a quantum dot in a two-dimensional hole gas, situated in a narrow-channel field-effect transistor in intrinsic silicon. Two layers of aluminum gate electrodes are defined on Si/SiO$_2$ using electron-beam lithography. Fabrication and subsequent electrical characterization of different devices yield reproducible results, such as typical MOSFET turn-on and pinch-off characteristics. Additionally, linear transport measurements at 4 K result in regularly spaced Coulomb oscillations, corresponding to single-hole tunneling through individual Coulomb islands. These Coulomb peaks are visible over a broad range in gate voltage, indicating very stable device operation. Energy spectroscopy measurements show closed Coulomb diamonds with single-hole charging energies of 5--10 meV, and lines of increased conductance as a result of resonant tunneling through additional available hole states.
\end{abstract}

\maketitle

In order for sufficient coherent operations to be performed in a proposed quantum computer \cite{LaddNature10}, the quantum states of the corresponding qubits are required to be long-lived. In the scheme proposed by \textcite{LossVin}, quantum logic gates perform operations on coupled spin states of single electrons in neighboring quantum dots.  Most experiments have focused on quantum dots formed in III-V semiconductors, especially GaAs \cite{WilfredDQDRMP,Hanson}; however, electron spin coherence in those materials is limited by hyperfine interactions with nuclear spins and spin-orbit coupling. Group IV materials are believed to have long spin lifetimes because of weak spin-orbit interactions and the predominance of spin-zero nuclei. This prospect has stimulated significant experimental effort to isolate single charges in carbon nanotubes \cite{jarillo, kuemmeth}, Si/SiGe heterostructures \cite{simmonsAPL07, borselliAPL2011}, Si nanowires \cite{zwanenburgNL09}, planar Si MOS structures \cite{lim}, and dopants in Si \cite{lansbergen, fuechsle, prati}. Silicon not only holds promise for very long coherence times \cite{stegerScience12}, but also for bringing scalability of quantum devices one step closer, and has thus attracted much attention for quantum computing purposes \cite{mortonNature11, zwanenburg-condmat}.

Recently, coherent driven oscillations of individual electron and nuclear spins in silicon were reported \cite{pla2012, pla2013}. The spin resonance was magnetically driven by sending alternating currents through a nearby microwave line. A technologically more attractive way is electric-field induced electron spin resonance, as demonstrated in quantum dots made in GaAs/AlGaAs heterostructures \cite{nowack, laird, pioro}, InAs nanowires \cite{nadjperge}, and  InSb nanowires \cite{pribiag}. Electrical control of single spins requires mediation by either hyperfine or spin-orbit interaction. Although the latter is too weak for electrically driven spin resonance of electrons in silicon, the spin-orbit interaction for holes may well facilitate hole spin resonance by means of electric fields.

Up until now, single-hole spins have not yet been investigated in electrostatically defined silicon quantum dots. Here, we report on single-hole tunneling (SHT) in a gated silicon MOSFET nanostructure, based on an earlier n-type design by \textcite{angus}. In this work we focus on low-temperature transport measurements through a two-dimensional hole gas (2DHG), which is electrostatically defined by a MOSFET-type architecture. To create the 2DHG we apply a negative potential to metallic gates on top of oxidized intrinsic silicon, raising the valence band to above the Fermi energy, thus allowing states to be occupied by holes. At 4 K we observe single-hole tunneling and demonstrate control of the charge occupation in unintentionally created quantum dots. We are aware that similar results have simultaneously been obtained elsewhere \cite{Li2013}.

\begin{figure}
\subfigure{

		\includegraphics[width=80mm]{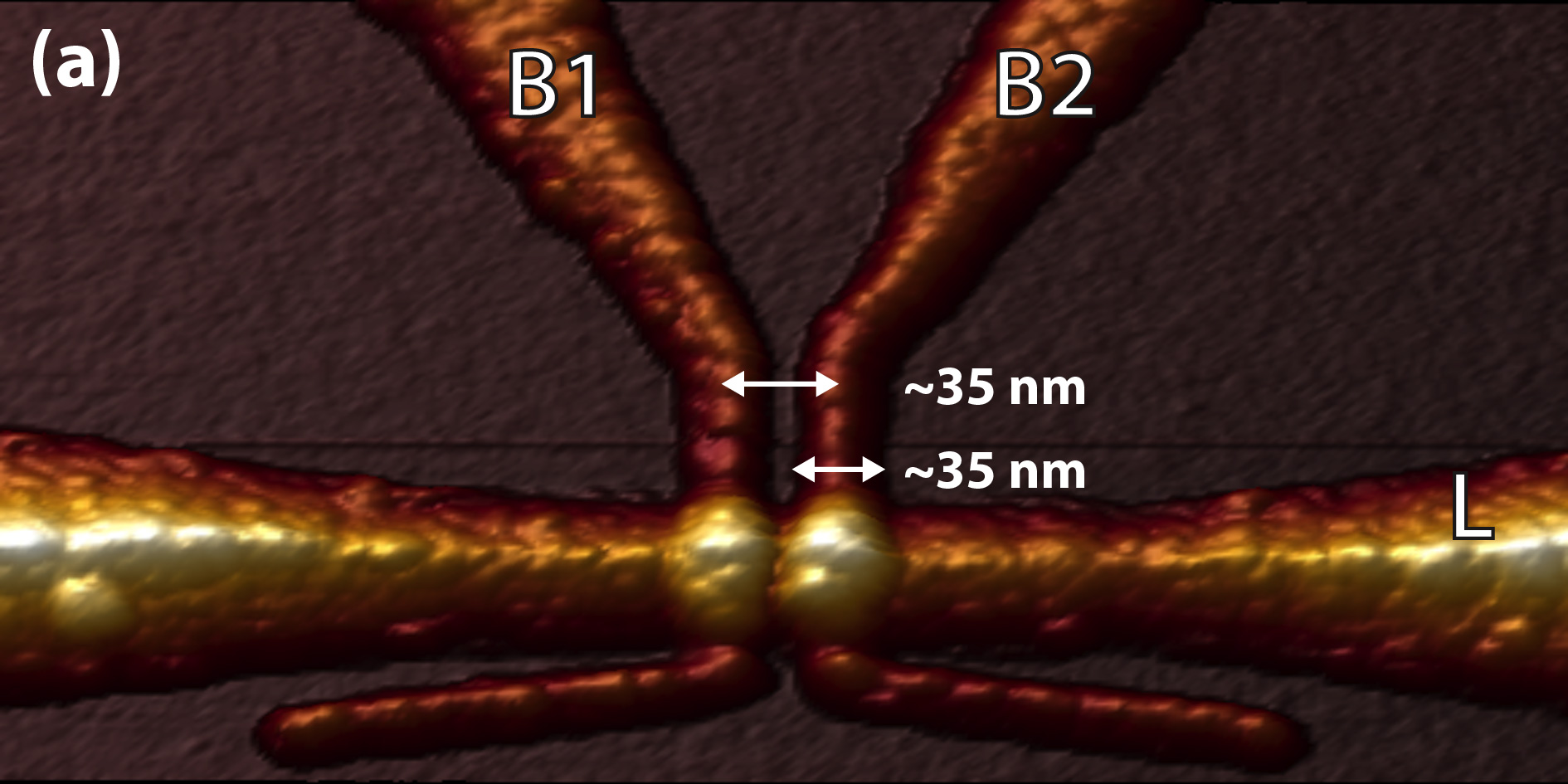}
		\label{fig:SEM}
}
\subfigure{
		\includegraphics[width=80mm]{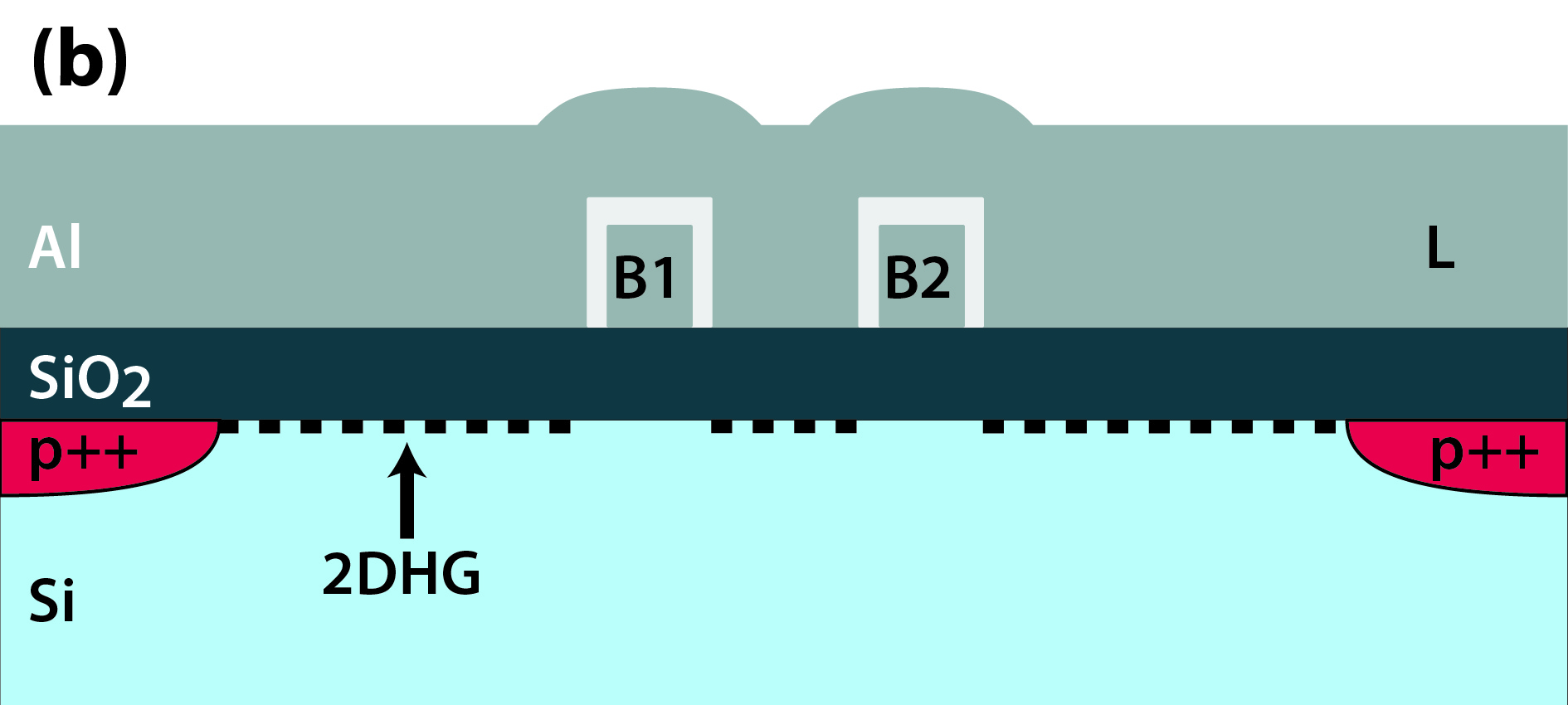}
		\label{fig:schematic}
}
	\caption{\textbf{Si quantum dot gate structure.} \subref{fig:SEM} Atomic force microscopy image of the device, showing the lead gate L horizontally across the image. The barrier gates B1 and B2 come in from the top center of the image. \subref{fig:schematic} A schematic cross-sectional image of the device. The highly p-type doped source and drain regions are shown in the intrinsic silicon, on top of which is the SiO$_2$ barrier. The Al gates are evaporated on top and electrically isolated from each other by aluminum oxide. The applied voltage on the aluminum gates creates a 2DHG, indicated by the dashed lines.}
	\label{fig:one}
	
\end{figure}

Figure~\ref{fig:one} shows an atomic force microscope image and a schematic cross section of the device structure, made with a combination of optical and electron-beam lithography (EBL), based on the recipe as described by \textcite{angus}. Near-intrinsic silicon ($\rho \geq 10000~\Omega\text{-}\text{cm}$) is used as the substrate. Source and drain regions are implanted with boron dopant atoms, which are activated by rapid thermal annealing, and serve as hole reservoirs. Ohmic contacts to these regions are made by sputtering  Al-Si alloy (99:1) contact pads. A $10$ nm thick high-quality SiO$_2$ oxide window is thermally grown at $900^\circ\text{C}$ and serves as an insulating barrier between the substrate and the aluminum gates. To remove charge traps and defects such as dangling bonds, the oxide is annealed in pure $\text{H}_{2}$ at $400^\circ\text{C}$ and a pressure of $10$ mbar. Contact pads for gates are defined using optical lithography followed by development, evaporation of Ti/Pt, and subsequent lift-off. EBL is used to define the sub-micron aluminum gates, which will electrostatically control hole accumulation.
Atomic force microscopy images show barrier gates with a typical width, height, and separation of $35$ nm, see figure~\ref{fig:SEM}. After oxidation of the barrier gates, a second EBL step is used to define the lead gate.
We have measured various devices and report here characteristic behavior of a representative sample. 

\begin{figure}
	\includegraphics[width=80mm]{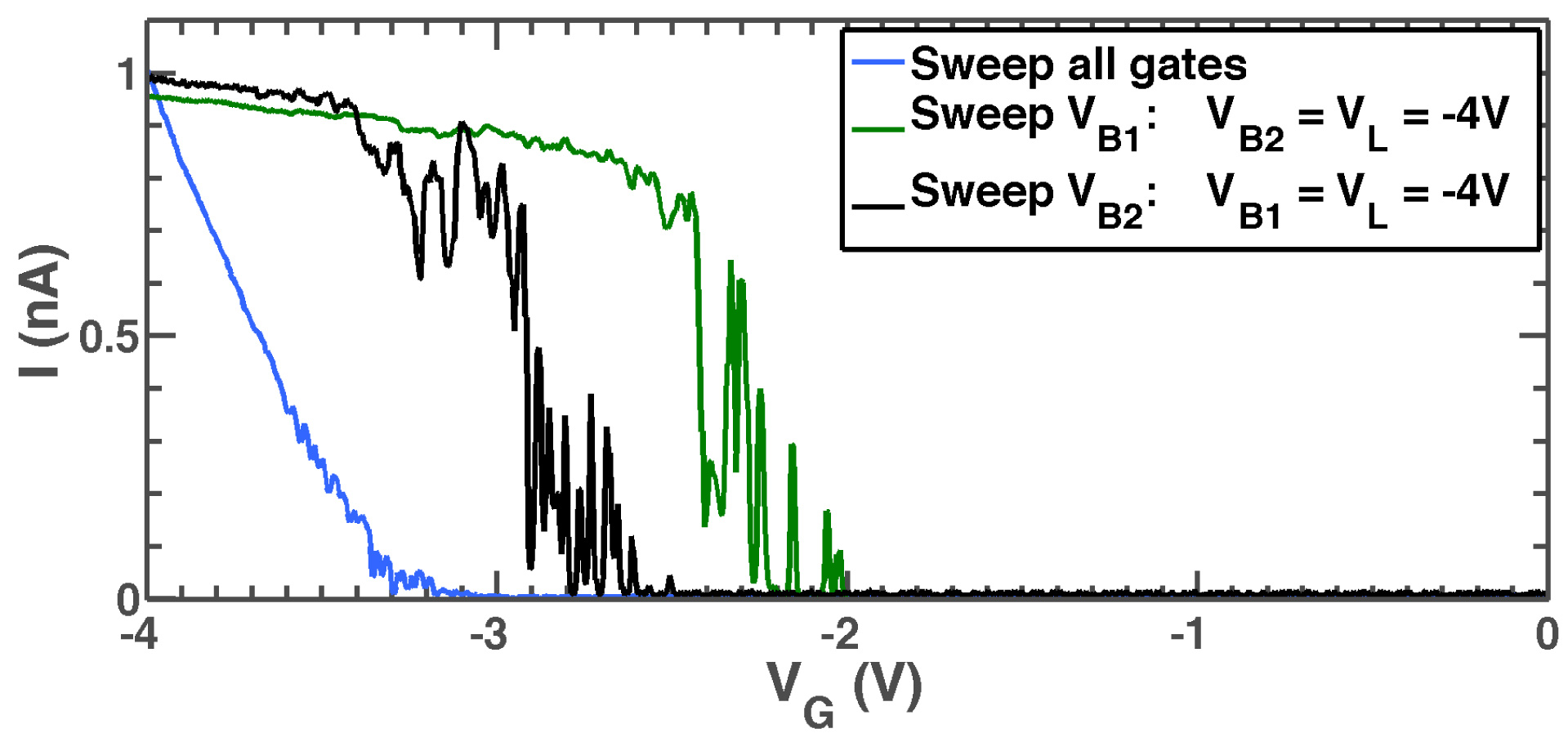}
	\caption{\textbf{MOSFET type turn-on and pinch-off behavior at $\mathbf{T \approx 4.2}$ K.} A bias voltage of $V_{SD} = 1~\text{mV}$ is applied between source and drain contacts. For turn-on, one single voltage is applied to all the gates and increased. Pinch-off curves are measured by setting the voltage equal on all gates except the pinch-off barrier. The curves for turn-on, pinch-off with B1, and pinch-off with B2 are blue, green, and black, respectively.}
		\label{fig:turnon}
\end{figure}

To characterize hole transport in our devices we perform electrical transport measurements on samples submerged in liquid helium at a temperature of $T \approx 4.2$ K. Low-noise current amplifiers and voltage sources in combination with Pi-filters were used to characterize the devices.
To measure a typical MOSFET turn-on characteristic, the same voltage $V_G$ is applied to all gates: $V_{L} = V_{B1} = V_{B2}$. Simultaneously, a bias voltage $V_{SD}$ is applied to the source and drain contacts. The gate voltages are then ramped to negative voltages while measuring the resulting current $I_{SD}$. Once the threshold voltage $V_{\text{Th}}$ is reached, the valence band is pulled sufficiently above the Fermi energy so that hole states become available to be occupied. In the resulting 2DHG, holes can then flow from source to drain and the device is `turned on'. During the sweeping of the gate voltages, the current increases up to roughly 1 nA, as shown in figure~\ref{fig:turnon}.
The ability of the barrier gates to influence conduction in the 2DHG below the barriers is critical to operation of the devices. The barrier gate voltages $V_{B1}$ and $V_{B2}$  should be able to tune the corresponding potential barriers from highly transparent ($I_{SD} \geq 1$ nA) to opaque ($I \approx 0$ nA). To test this, a `pinch-off' curve is measured, by making the voltage on a barrier less negative whilst keeping the other gates well beyond the threshold voltage. Both barriers B1 and B2 can individually pinch off the conduction channel, and additionally show some resonances in the measured current. The results in Figure~\ref{fig:turnon} show the ability of each of the barriers to effectively control conduction in the channel.

\begin{figure}
\subfigure{
		\includegraphics[width=80mm]{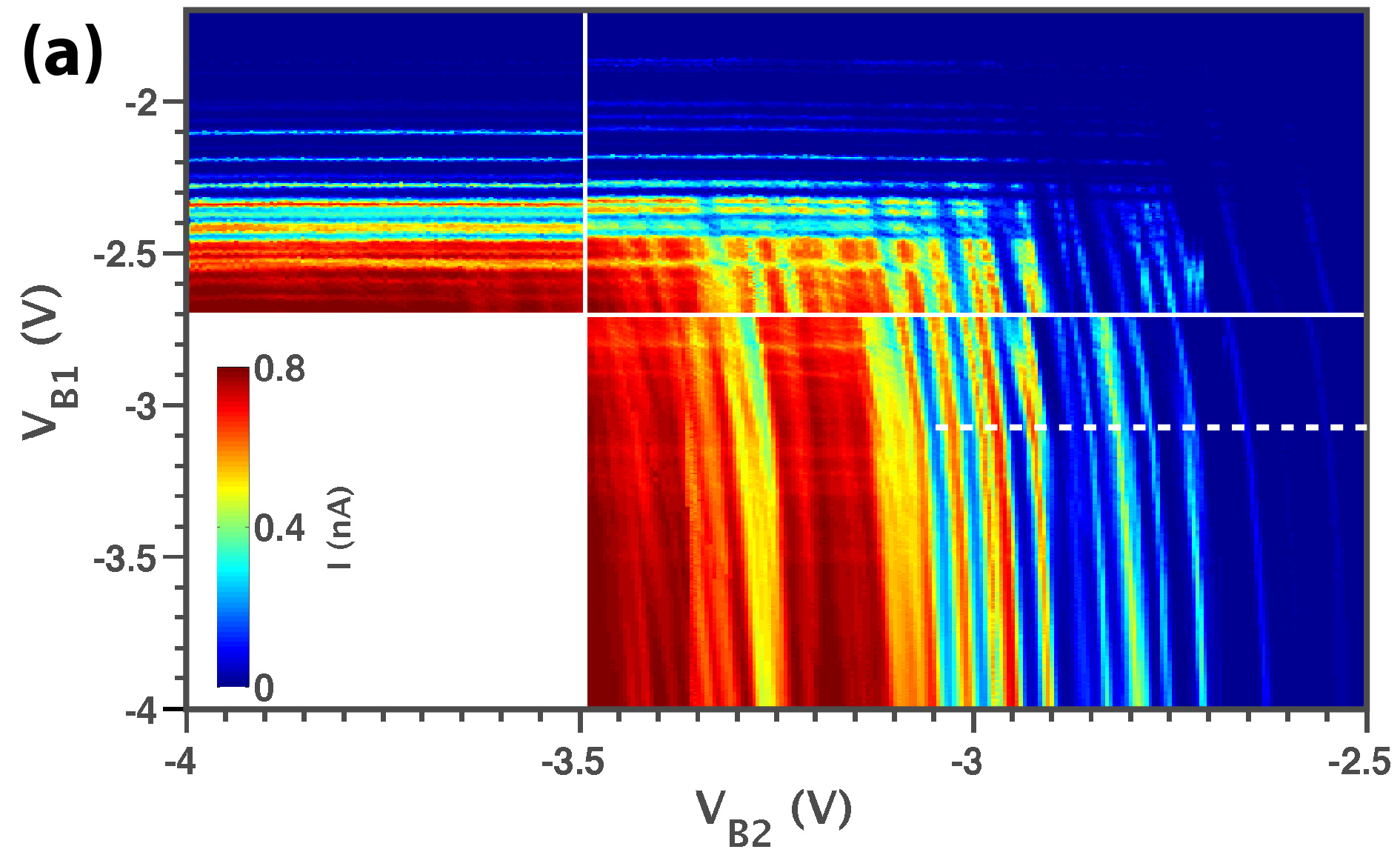}
		\label{fig:b1b2}
	}
\subfigure{
		\includegraphics[width=80mm]{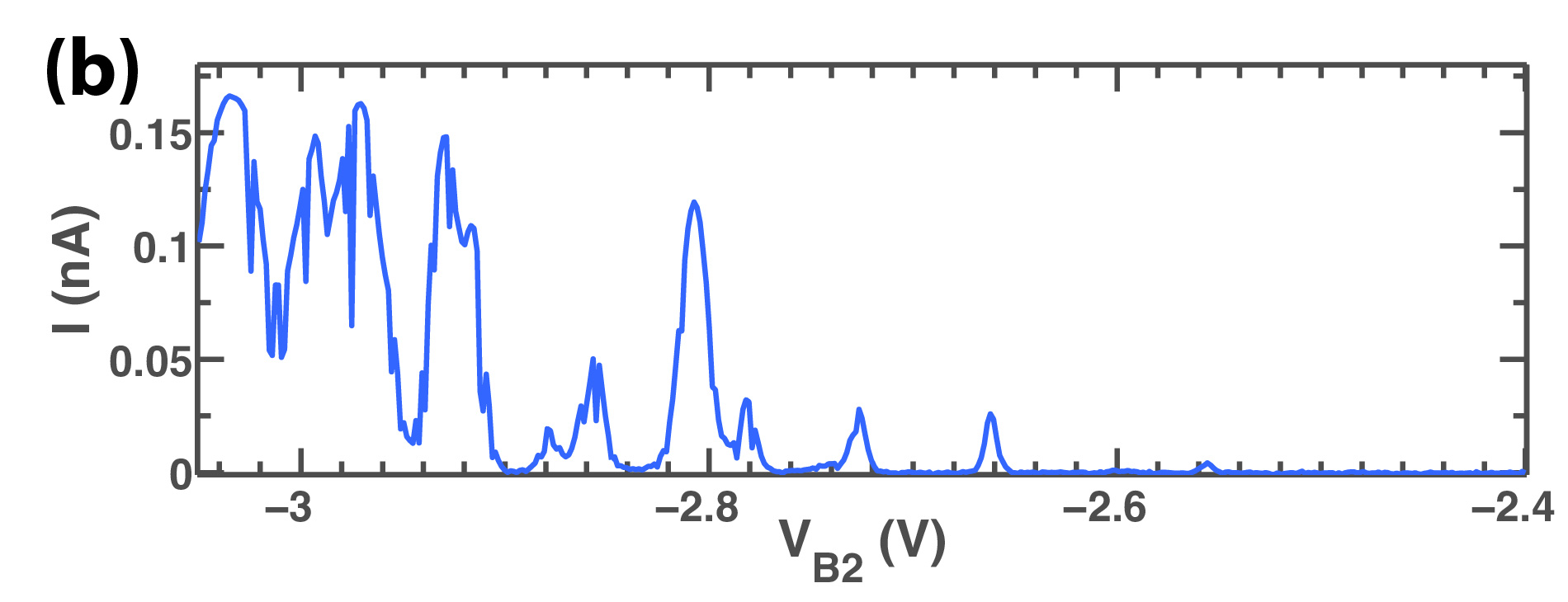}
		\label{fig:coulombpeaks}
	}
	\caption{\textbf{Single-hole tunneling in the linear transport regime.} \subref{fig:b1b2} The current as a function of applied barrier voltages $V_{B_1}$, $V_{B_2}$ with $V_{SD} = 1~\text{mV}$, showing periodicities in several directions. Three measurements at different times are visible, hence the discontinuity at $V_{B1} = -2.7~\text{V}$ and $V_{B2} = -3.5~\text{V}$.
	\subref{fig:coulombpeaks} Coulomb peaks in the current with varying $V_{B2}$, taken at the dashed line in \subref{fig:b1b2} at constant $V_{SD} = 0.3~\text{mV}$, $V_L = -3.95~\text{V}$, and $V_{B1} = -3.1~\text{V}$
	}
	\label{fig:three}	
\end{figure}

Next, we measure the current versus both barrier gate voltages at constant bias and lead gate voltage (see Figure~\ref{fig:b1b2}). When the voltage applied to the barrier gates is too close to zero ($V_{B1} \geq -2~\text{V}$ or $V_{B2} \geq -2.6~\text{V}$), the tunneling rate through the potential barriers becomes negligible.
The current as a function of voltage applied to the barrier gates shows periodic resonances parallel to the axes of both $V_{B1}$ and $V_{B2}$ . Resonances parallel to the axis of $V_{B1}$ ($V_{B2}$) are not influenced by a change in $V_{B2}$ ($V_{B1}$), indicating that those features are independently coupled to $V_{B1}$ and $V_{B2}$. We now focus on one of these resonances; specifically the one coupled most strongly to B2. To probe the features of this resonance, a constant $V_{SD}$ and $V_{B1}$ is applied where $V_{B1}$ is chosen such that the corresponding barrier is highly transparent. The subsequent measured source-drain current as a function of $V_{B2}$ in figure~\ref{fig:coulombpeaks} shows periodic current peaks separated by regions of zero current. The sharp peaks correspond to Coulomb oscillations with regions of Coulomb blockade in between. Each time a peak is traversed, the occupation of the corresponding Coulomb island changes by one hole in a charge transition $N \leftrightarrow N+1$, with $N$ the number holes on the island. We can thus control the charge occupation of individual islands below the barrier gates. These islands are likely formed by disorder or roughness, e.g. impurities or charge traps in the SiO$_2$. We conclude that figure~\ref{fig:coulombpeaks} shows the trademark of single-hole tunneling and control of charge occupation in intrinsic silicon.

\begin{figure}[!t]
	\includegraphics[width=80mm]{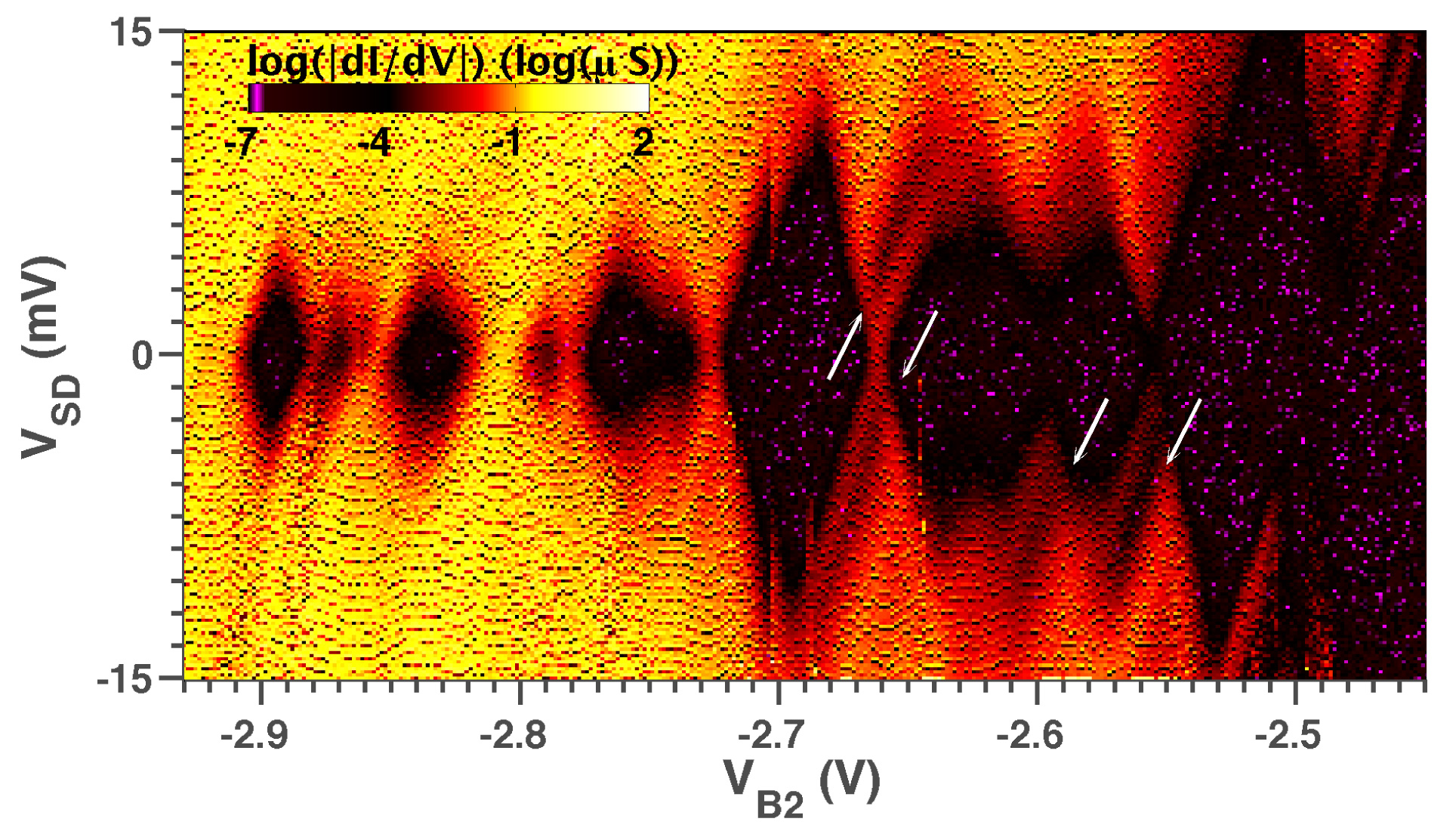}
	\caption{\textbf{Single-hole tunneling in the non-linear transport regime.} Bias spectroscopy taken at $V_{B_1}=-3.1~\text{V}$ and $V_L = -3.95~\text{V}$. Resonant tunneling features are visible and are indicated by arrows.}
	\label{fig:biasspectroscopy}
	
\end{figure}

Energy spectroscopy was used to further characterize the device. The numerical differential conductance $dI/dV_{SD}$ with varying $V_{B2}$ and $V_{SD}$ is suppressed periodically by Coulomb blockade, as shown in figure~\ref{fig:biasspectroscopy}. The Coulomb diamonds are reasonably well defined and exhibit a single period, as evidenced by the parallel peaks in Fig.~\ref{fig:three}. The shape of the diamonds is most likely modulated by fluctuations in the conductance elsewhere in the device and the electrostatic environment, e.g. charge traps. At the diamond edges the electrochemical potential of the corresponding Coulomb island is resonant with either source or drain potential and single holes tunnel through the device. Most Coulomb diamonds close at zero source-drain bias, again indicating transport through a single island. $V_{B2}$ changes the charge occupation of the island from $N$ to $N\pm1$ by moving from one Coulomb diamond to the next .
The Coulomb diamonds have very similar shapes across a wide voltage range and reproduce in repeated measurements, indicating the robustness of the device.



The charging energy $E_C$ of the island varies from $\sim$~5 to 10 meV. This corresponds to an island capacitance of $C \approx$ 32 to 16 aF, which corresponds to a diameter of the island of about 76 to 38 nm in a classical disc capacitor model -- where $E_c = e^2 / 4 \epsilon_0 \epsilon_{\text{Si}} d$, with $d$ the diameter of the island.
In the last few diamonds, lines of increased conductance appear parallel to the diamond edges at positive and negative bias. We attribute this to resonant tunneling features as a results of extra available states for tunneling lined up with either source or drain. These features may correspond to orbital excited states, although the origin cannot be determined based on these data alone \cite{EscottN10}. The results in Fig.~\ref{fig:biasspectroscopy} show single-hole tunneling probed by energy spectroscopy, in which the present resonant tunneling features underline the ability to observe quantum states in these single-hole transistors.

To conclude, we have reported the fabrication and electronic characterization of p-type narrow-channel field-effect transistors in intrinsic silicon. Aluminum gate structures made on Si/SiO$_2$ with electron-beam lithography were used to create and control a two-dimensional hole gas at the interface of silicon and silicon oxide. Hole transport at 4 K can be controlled by barrier and lead gates, such that Coulomb peaks appear at small source-drain bias. Highly regular Coulomb peak lines and closing Coulomb diamonds in energy spectroscopy clearly indicate single-hole tunneling in the many-hole regime. The strong capacitive coupling to each respective barrier gate suggest that single Coulomb islands are created underneath or in the vicinity of the controlling gate. These islands are caused by local potential fluctuations due to e.g. impurities or charge traps in the SiO$_2$ or at the interface of Si and SiO$_2$. Silicon is known for being extremely sensitive to disorder, owing to the large effective mass of the charge carriers, which is even higher for holes than for electrons. The evidence for resonant tunneling features in energy spectroscopy indicates that these devices have demonstrable quantum confinement, even at relatively high temperatures. Further optimization of the fabrication process will focus on (i) improvement the material quality of metal, oxide and semiconductor, e.g. lowering the charge trap density in the SiO$_2$ and reducing the grain size in the Al, and (ii) minimization of disorder and roughness at the material interfaces, e.g. removing dangling bonds by introducing extra annealing steps. The resulting low-disorder hole quantum dots with tunable tunnel barriers pave the way towards control of single holes and single spins in silicon.





%

\end{document}